\documentclass[10pt,conference,a4paper]{IEEEtran}
\usepackage{epsfig,graphicx,psfrag,amsmath,cases,bm}
\usepackage{latexsym,amssymb,algorithm,mathtools}
\usepackage{algorithmic}
\usepackage{color}
\usepackage{url}
\usepackage{scrtime}
\usepackage{stfloats}
\usepackage{tablefootnote}
\usepackage{cite}
\usepackage{amsfonts,mathabx}
\usepackage{textcomp}
\usepackage{amsthm}
\usepackage{geometry}
\usepackage{enumitem}
\usepackage{titlesec}
\usepackage{siunitx} 
\allowdisplaybreaks
\ifCLASSOPTIONcompsoc
    \usepackage[caption=false, font=normalsize, labelfont=sf, textfont=sf]{subfig}
\else
\usepackage[caption=false, font=footnotesize]{subfig}
\fi
\renewcommand{\thesubsubsection}{\arabic{subsubsection}}

\titleformat{\subsubsection}[runin]{\itshape}{\thesubsubsection)}{1em}{}
\titlespacing*{\subsubsection}{\parindent}{0pt}{*1}
\titlespacing*{\section}{0pt}{0.05\baselineskip}{0.05\baselineskip}
\titlespacing*{\subsection}{0pt}{0.05\baselineskip}{0.05\baselineskip}

 \def\BibTeX{{\rm B\kern-.05em{\sc i\kern-.025em b}\kern-.08em
    T\kern-.1667em\lower.7ex\hbox{E}\kern-.125emX}}
    
\usepackage{xcolor,capt-of}
\usepackage{multirow}
\makeatletter
\newcommand*{\rom}[1]{\expandafter\@slowromancap\romannumeral #1@}
\makeatother

\DeclareMathOperator{\mino}{minimize}

\makeatletter
\def\thm@space@setup{\thm@preskip=0pt
\thm@postskip=1pt}
\makeatother

\IEEEaftertitletext{\vspace{-3mm}}
\IEEEaftertitletext{\vspace{-2.6em}}

\title{Movable-Antenna-Enhanced ISAC: Optimal Antenna Trajectory and Beamforming Design\vspace*{-4mm}}
\author{\vspace*{-0.3cm}\IEEEauthorblockN {Yifei Wu\IEEEauthorrefmark{1}, Dongfang Xu\IEEEauthorrefmark{2}, Derrick Wing Kwan Ng\IEEEauthorrefmark{3}, Wolfgang Gerstacker\IEEEauthorrefmark{1}, and Robert Schober\IEEEauthorrefmark{1}}\\
\IEEEauthorrefmark {1}Friedrich-Alexander-Universit\"at, Germany;
\IEEEauthorrefmark {3}University of New South Wales, Australia\\
\IEEEauthorrefmark {2}The Hong Kong University of Science and Technology, Hong Kong 
\vspace*{-0.5cm}}

\begin{document}

\maketitle
\vspace{-1.5cm}
\begin{abstract}
Integrated sensing and communication (ISAC) is a key enabling technology for next-generation wireless networks. However, most existing ISAC systems rely on fixed-position antennas, which restrict performance when balancing sensing and communication objectives. Movable antenna (MA) technology introduces additional spatial degrees of freedom through antenna mobility, yet existing studies on MA-enabled ISAC schemes mainly consider static antenna repositioning and fail to fully exploit this capability. By leveraging spatio-temporal sampling enabled by antenna motion, optimized MA trajectories can synthesize large virtual aperture arrays, thereby improving angular resolution and reducing sensing ambiguity. To this end, this paper investigates a dynamic MA-enabled ISAC system and studies the joint design of MA trajectories and transmit beamforming. We formulate a joint trajectory and beamforming optimization problem to minimize sensing beampattern mismatch under communication quality-of-service constraints.
A branch-and-bound–based algorithm is developed to obtain the globally optimum solution. Numerical results show that the proposed framework significantly outperforms baseline schemes with only one or two antenna repositioning steps, demonstrating its practical feasibility.
\end{abstract}
\section{Introduction}
Integrated sensing and communication (ISAC) has emerged as a key paradigm for alleviating the radio spectrum scarcity problem in upcoming sixth-generation (6G) wireless networks \cite{wong2017key}. In particular, by employing novel dual-functional radar-communication (DFRC) base
stations (BSs), ISAC enables the simultaneous support of high-rate data transmission and high-resolution environmental sensing via shared spectrum and hardware \cite{liu2020joint}. By seamlessly integrating radar-like sensing and communication functionalities, ISAC is expected to play a pivotal role in enabling an emerging wide range of applications such as autonomous driving, extended reality, smart manufacturing, and human–machine interaction \cite{wei2022toward}. However, realizing the full potential of ISAC typically relies on leveraging large-scale multiple-input and multiple-output (MIMO) arrays at both the transmitter and receiver sides, which introduces new challenges in system design, including the deployment, calibration, and real-time control of dense antenna arrays, as well as the associated increase in signal processing complexity and computational burden \cite{xu2022robust}.

To exploit the spatial degrees of freedom (DoFs) inherent to large-scale MIMO systems, recent advances in antenna technology have given rise to the concept of movable antennas (MAs), whose physical locations can be dynamically adjusted within a predefined region. Unlike conventional fixed antennas, MA elements can be repositioned in response to time-varying and heterogeneous channel conditions by employing electromechanical actuators such as stepper motors and servos \cite{zhuravlev2015experimental,zhu2023movable_chanllenges}. This capability enables the system to establish controllable and favorable spatial antenna correlations, thereby significantly enhancing spectral efficiency and sensing performance. While MAs have recently attracted significant attention for wireless communication system design, their potential in ISAC systems remains largely unexplored. Some initial works, e.g., \cite{khalili2024advanced,chen2025antenna} have investigated joint MA position and beamforming design to optimize ISAC performance. However, these works primarily demonstrated performance improvements due to static antenna repositioning, without revealing the fundamental relationship between the MA spatial configuration and sensing performance \cite{ma2025movable}.  
In fact, such a static MA-enhanced ISAC scheme cannot fully exploit the mobility provided by MA elements. In particular, dynamically adjusting MA positions introduces additional spatial sampling points across sensing snapshots. The measurements collected for these time-varying antenna locations can be coherently combined, thereby forming a large virtual aperture array synthesized over time at the transmitter\cite{comesana2014introduction}. 
Therefore, by optimizing the MA trajectory in ISAC systems, we can construct a large-scale spatio-temporal virtual aperture array with reduced the correlation among steering vectors associated with different angular directions, mitigating ambiguity and interference in angle estimation. 
Motivated by these observations, this paper investigates an MA-enhanced ISAC framework and studies the joint design of MA trajectories and beamforming to optimize both communication and sensing performance. We consider a dynamic MA-ISAC scenario, where MA elements can be repositioned frequently and continuously within a prescribed region, and aim at exploiting this mobility to improve target sensing performance while guaranteeing a minimum quality of service (QoS) requirement for all communication users. 
The proposed approach provides new physical-layer insights into trajectory-enabled virtual-aperture synthesis, establishing MA spatial mobility as a new design resource for next-generation ISAC systems. 

The remainder of this paper is organized as follows. Section II introduces the system and signal models for the MA-enhanced ISAC system. Section III presents the resource allocation problem formulation, and Section IV develops a globally optimal joint antenna trajectory and beamforming design algorithm. Section V provides numerical performance results. Finally, Section VI concludes the paper.

\textit{Notation:} 
Vectors and matrices are denoted by boldface lower-case and boldface capital letters, respectively. $\mathbb{R}^{N\times M}$ and $\mathbb{C}^{N\times M}$ denote the space of $N\times M$ real-valued and complex-valued matrices, respectively. $||\cdot||_2$ and $||\cdot||_{\infty}$ denote the $l_2$-norm and L-infinity norm of the argument, respectively. $||\cdot||_{\mathrm{row-0}}$ and $||\cdot||_{\mathrm{col-0}}$ denote the 
number of non-zero rows and columns of a matrix, respectively. $(\cdot)^T$, and $(\cdot)^H$ stand for the transpose, and conjugate transpose of their input arguments, respectively. $\mathbf{I}_{N}$ refers to the identity matrix of dimension $N$. $\mathrm{Tr}(\cdot)$ and $\mathrm{Rank}(\cdot)$ denote the trace and rank of the input matrix. $\mathbf{X}(i,:)$ and $\mathbf{X}(:,i)$ denotes
the $i$-th row vector and column vector of matrix $\mathbf{X}$, respectively. 
$\lfloor\cdot\rfloor$ and $\lceil\cdot\rceil$ denote the floor and ceil functions, respectively. $m \mod n$ represents the the remainder of the division of $m$ by $n$.
\section{System Model}
\subsection{MA-enhanced ISAC System Model}
\begin{figure}
    \centering
    \includegraphics[width=3.0 in]{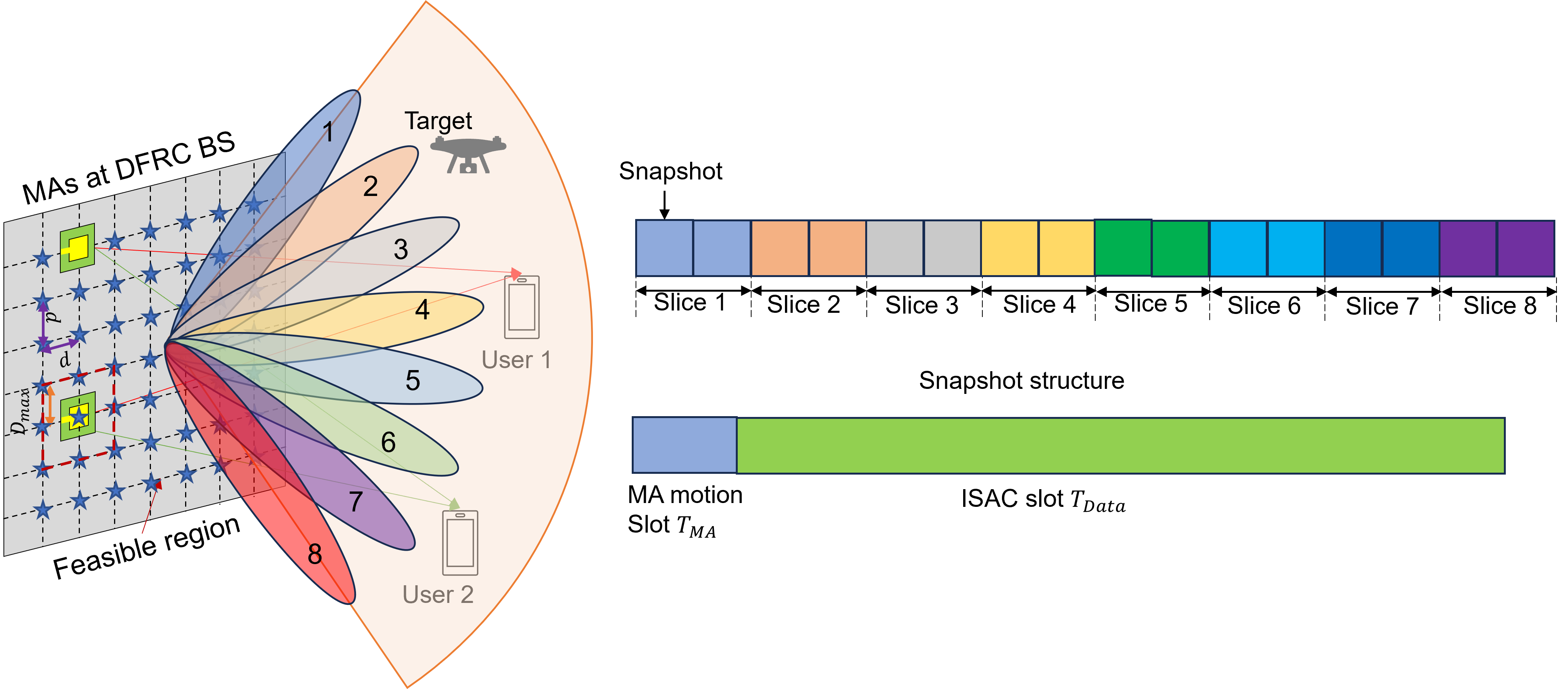}
    \caption{Considered system model and frame structure consisting of $M=2$ MA elements, $K=2$ users, $Q=8$ slices, $N=2$ snapshots each slice.}\vspace{1mm}
    \label{fig:placeholder}
\end{figure}
As shown in Fig. 1, we consider a downlink multiuser bistatic ISAC system. 
Specifically, a 
DFRC BS equipped with $M$ MA elements scans a sector area exploiting a sequence of $Q$ ISAC MA beams to detect new targets and update sensing information within a time period $T_{\mathrm{tot}}$\cite{xu2022robust}. 
The sensing sector area is divided into $Q$ angular slices, and each ISAC MA beam is steered toward a specific direction so as to illuminate one angular slice with a desired directional power gain. 
Moreover, we adopt a virtual antenna aperture technique exploiting dynamic MA repositioning. We allocate $N$ sensing snapshots per angular slice, during which the BS dynamically adjusts the MA positions and transmits ISAC signals. This process provides $N\times M$ spatio-temporal sampling points for sensing, as it naturally synthesizes a larger virtual phased array with $N\times M$ virtual antenna elements. 
Then, the sensing receiver collects the received ISAC signal over the $N$ snapshots and applies an optimal linear combining strategy to fully realize the sensing gain provided by the virtual antenna array. Meanwhile, during these $N$ sensing snapshots, the BS must guarantee the QoS requirement of $K$ single-antenna users by jointly designing the MA positions and beamforming. 
\subsection{MA Model}
Since practical electromechanical devices can only provide horizontal or vertical displacement with a fixed step size $d$ \cite{zhuravlev2015experimental}, the transmitter region of the MA-enhanced communication system is spatially quantized. We collect the $J$ possible discrete positions of the MAs into the set $\mathcal{P}=\{\mathbf{p}_1,\cdots, \mathbf{p}_J\}$, where the distance between neighboring positions is equal to $d$ in horizontal or vertical direction, as shown in Fig. 1, and $\mathbf{p}_j=[x_j,y_j]^T$ represents the $j$-th candidate position with horizontal coordinate $x_j$ and vertical coordinate $y_j$. On the other hand, due to the finite speed of the MA motion drivers, the displacement of the MA elements in one snapshot is limited. We assume that the horizontal and vertical velocities of MAs are the same and constant, i.e.,   $v_{v,\mathrm{MA}}=v_{h,\mathrm{MA}}=v_{\mathrm{MA}}$. Accordingly, for one snapshot, the maximum vertical and horizontal displacement of each MA element during the motion interval is given by $D_{\mathrm{max}}=v_{\mathrm{MA}}T_{\mathrm{MA}}$, where $T_{\mathrm{MA}}$ is the MA motion time
Therefore, in snapshot $n$, the position $\mathbf{t}_m[n]=[{t}_{m,x}[n],{t}_{m,y}[n]]^T$ of MA element $m$ must satisfy the inter-snapshot mobility constraint\cite{wu2025globallytcom}, i.e., $|{t}_{m,x}[n]-{t}_{m,x}[n-1]|\leq D_{\mathrm{max}}$ and $|{t}_{m,y}[n]-{t}_{m,y}[n-1]|\leq D_{\mathrm{max}}$, which can be recast as
\begin{equation}\label{Max_dis_cons}
    \|\mathbf{t}_m[n]-\mathbf{t}_m[n-1]\|_{\infty}\leq D_{\mathrm{max}},
\end{equation}
where $\mathbf{t}_m[n-1]=[\mathbf{t}_{m,x}[n-1],\mathbf{t}_{m,y}[n-1]]^T$ denotes the position of MA element $m$ in snapshot $n-1$. Due to physical antenna-size constraints, two MA elements cannot be placed arbitrarily close to each other \cite{zhu2022modeling}. Thus, in all snapshots, the center-to-center distance between any pair of MA elements needs to be greater than a minimum distance $D_{\mathrm{min}}$:
\begin{equation}\label{Min_dis_cons}
    \|\mathbf{t}_m[n]-\mathbf{t}_{m'}[n]\|_2\geq D_{\mathrm{min}},\ m\neq m'
\end{equation}
\subsection{Signal Model}
Without loss of generality, we focus on the ISAC design for sensing in the first angular slice\footnote{The proposed method can be easily extended for the design of sensing for the other slices by varying the beam direction. }. In snapshot $n$, the BS transmits a ISAC waveform $\mathbf{x}[n]\in\mathbb{C}^{M}$, which is given by
\begin{equation}
   \mathbf{x}[n]=\mathbf{W}[n]\mathbf{s}[n]+\mathbf{v}[n], 
\end{equation}
where $\mathbf{W}[n]=[\mathbf{w}_{1}[n],\dots,\mathbf{w}_{K}[n]]\in\mathbb{C}^{M\times K},\mathbf{s}[n]=[s_1[n],\cdots,s_{K}[n]]^T,$ and $\mathbf{v}[n]\in\mathbb{C}^{M}$ denote the beamforming matrix, transmitted information symbol vector, and sensing radar signal vector in time slot $n$, respectively. Here, $\mathbf{w}_k[n]\in\mathbb{C}^{M}$ and $s_k[n]\in\mathbb{C}$ denote the beamforming vector and the transmitted information symbol of user $k$ in snapshot $n$, respectively. Moreover, the covariance matrix of vector $\mathbf{s}[n]$ is given by $\mathbb{E}\{\mathbf{s}[n]\mathbf{s}^H[n]\}=\sigma_s^2\mathbf{I}_{K}$, where $\sigma_s^2$ denotes the variance of the transmit symbols. To fully utilize the sensing gain provided by MAs, the BS designs the joint covariance matrix $\mathbf{R}$ of the sensing radar signals for all $N$ snapshots. Let vector $\mathbf{v}=[\mathbf{v}^H[1],\cdots, \mathbf{v}^H[N]]^H$ collect the sensing radar signals of all $N$ snapshots. Then, the covariance matrix $\mathbf{R}$ is defined as $\mathbf{R}=\mathbb{E}\{\mathbf{v}\mathbf{v}^H\}\succeq\mathbf{0}$.
In MA systems, the physical channel between the MA elements and users can be reconfigured by adjusting the MA positions. Here, the channel coefficient between MA element $m$ and user $k$ in time slot $n$ is denoted as $h_k(\mathbf{t}_m[n])\in\mathbb{C}$. 
Then, the received signal ${y}_k[n]$ at user $k$ in time slot $n$ is given by
\begin{equation}
    {y}_k[n]=\mathbf{h}_k^H(\mathbf{t}[n])\mathbf{x}[n]+{n}_k,
\end{equation}
where $\mathbf{h}_k(\mathbf{t}[n])=[h_k(\mathbf{t}_1[n]),\cdots, h_k(\mathbf{t}_M[n])]^H$ represents the channel vector from the $M$ MA elements to user $k$ in snapshot $n$. Here, $\mathbf{t}[n]=[\mathbf{t}^T_1[n],\cdots, \mathbf{t}_M^T[n]]^T$ collects the MAs' positions in snapshot $n$. Moreover, $n_k$ represents the additive white Gaussian noise at user $k$ with zero mean and variance $\sigma_k^2$. Here, we define sets $\mathcal{K}\in\{1,\cdots,K\}$, $\mathcal{M}\in\{1,\cdots,M\}$, and $\mathcal{N}\in\{1,\cdots,N\}$ to collect the indices of the users, MA elements, and snapshots, respectively.

For the sensing functionality, the sensing receiver can collect the received signal over $N$ consecutive snapshots.
We assume that both the transmitter-target and target-receiver links are dominated by line-of-sight (LoS) propagation. Moreover, a far-field channel model is considered \cite{khalili2024advanced}. Then, in snapshot $n$, the received signal for a specific target with elevation angle of departure (AoD) $\alpha_t$ and azimuth AoD $\beta_t$ is given by 
\begin{equation}
    \mathbf{y}_s[n]=\zeta\mathbf{a}_r\mathbf{a}_t^H(\alpha_t,\beta_t,\mathbf{t}[n])\mathbf{x}[n]+\mathbf{n}_s[n],
\end{equation}
where $\mathbf{a}_r$ and $\mathbf{a}_t^H(\alpha_t,\beta_t,\mathbf{t}[n])$ denote the receive and transmit steering vectors, respectively, i.e.,
\begin{equation}
    \mathbf{a}_t(\alpha_t,\beta_t,\mathbf{t}[n])=\left[e^{j\rho(\alpha_t,\beta_t,\mathbf{t}_1[n])},\cdots,e^{j\rho(\alpha_t,\beta_t,\mathbf{t}_M[n])}\right],\notag
\end{equation}
with $\rho(\alpha_t,\beta_t,\mathbf{t}_m[n])=\frac{2\pi}{\lambda}((x_m[n]-x_1)\cos\alpha_{t}\sin\beta_{t}+$ $(y_m[n]-y_1)\sin\alpha_{t})$ denoting the phase difference between the position of MA element $m$ in snapshot $n$, $\mathbf{t}_m[n]=[x_m[n],y_m[n]]^T$, and the reference position $\mathbf{p}_1=[x_1,y_1]^T$. 
Moreover, $\zeta$ and $\mathbf{n}_s[n]$ represent the complex channel coefficient and received noise in snapshot $n$, respectively. We note that due to the movement of the MA elements, the transmit steering vector varies across the snapshots. 
As a result, the collection of these snapshot-dependent steering vectors effectively constitutes a large virtual phased array synthesized over time\cite{comesana2014introduction,zhuravlev2015experimental}. 
According to \cite{comesana2014introduction}, to maximize the beam gain, the sensing receiver can adopt a delay-and-sum strategy to coherently combine the received signals over $N$ snapshots. Hence, the combined received signal $y_s$ is given by
    ${y}_s=\sum_{n\in\mathcal{N}}\mathbf{f}_r^H[n]\mathbf{y}_s[n],$
where $\mathbf{f}_r[n]$ denotes the linear combining vector at the sensing receiver in snapshot $n$. 
According to \cite{comesana2014introduction}, for the LoS target-receiver channel, a closed-form optimal solution for the linear combining vector in snapshot $n$ is given by $    \mathbf{f}_r^H[n]=\mathbf{a}^H_r(\alpha_r).
$
Then, we can rewrite ${y}_s$ as follows
\begin{equation}
\begin{aligned}
{y}_s&=N_R\zeta\sum_{n\in\mathcal{N}}\mathbf{a}_t^H(\alpha_t,\beta_t,\mathbf{t}[n])\mathbf{x}[n]+\sum_{n\in\mathcal{N}}\mathbf{f}_r^H[n]\mathbf{n}_s[n]\\[-3pt]
&=N_R\zeta\mathbf{a}_T^H(\alpha_t,\beta_t,\mathbf{t})\mathbf{x}_t+\sum_{n\in\mathcal{N}}\mathbf{f}_r^H[n]\mathbf{n}_s[n],\notag
\end{aligned}
\end{equation}
where $\mathbf{x}_t=[\mathbf{x}^H[1],\cdots,\mathbf{x}^H[N]]^H$ and $\mathbf{a}_T(\alpha_t,\beta_t,\mathbf{t})=[\mathbf{a}_t^H(\alpha_t,\beta_t,\mathbf{t}[1]),\cdots,\mathbf{a}_t^H(\alpha_t,\beta_t,\mathbf{t}[N])]^H$. Here, we define $\mathbf{t}=[\mathbf{t}^T[1],\cdots,\mathbf{t}^T[N] ]$ to collect the MA positions of all $N$ snapshots.
Here, maximizing the average received power is equivalent to maximizing $P(\alpha_t,\beta_t)$, i.e.,
\begin{equation}
\begin{aligned}
    &P(\alpha_t,\beta_t)=\mathbf{a}_T^H(\alpha_t,\beta_t,\mathbf{t})\mathbb{E}\{\mathbf{x}_t\mathbf{x}_t^H\}\mathbf{a}_T(\alpha_t,\beta_t,\mathbf{t})\\[-2pt]
    &=\sigma_s^2\sum_{n\in\mathcal{N}}\mathbf{a}_t^H(\alpha_t,\beta_t,\mathbf{t}[n])\mathbf{W}[n]\mathbf{W}^H[n]\mathbf{a}_t(\alpha_t,\beta_t,\mathbf{t}[n])\\[-2pt]
    &+\mathbf{a}_T^H(\alpha_t,\beta_t,\mathbf{t})\mathbf{R}\mathbf{a}_T(\alpha_t,\beta_t,\mathbf{t}).\notag
\end{aligned}
\end{equation}
Here, we define $ P(\alpha_t,\beta_t)$ as the sensing beam gain in elevation AoD $\alpha_t$ and azimuth AoD $\beta_t$. 
\section{Problem Formulation}
First, we define an effective channel vector $\hat{\mathbf{h}}_k=[h_k(\mathbf{p}_1),\cdots,h_k(\mathbf{p}_J)]^H$, which collects the channel coefficients from an MA element to user $k$ for all $J$ feasible discrete MA locations. Then, we can rewrite the channel coefficient ${h}_k(\mathbf{t}_m[n])$ from MA element $m$ to user $k$ in snapshot $n$ as ${h}_k(\mathbf{t}_m[n])=\mathbf{b}_m^T[n]\hat{\mathbf{h}}_k,$
where $\mathbf{b}_m[n]=\big[b_{m,1}[n],\cdots,b_{m,J}[n]\big]^T$ is a binary position-selection vector. Here, $b_{m,j}[n]\in\left\{0,1\right\}$, with the constraint $\sum_{j=1}^{J}b_{m,j}[n]=1$, is a binary variable defining the position of MA element $m$ in snapshot $n$. Then, we define binary matrix $\mathbf{B}[n]=[\mathbf{b}_1[n],\cdots,\mathbf{b}_M[n]]^T$ and rewrite channel vector $\mathbf{h}_k(\mathbf{t}[n])$ from the BS to user $k$ in snapshot $n$ as $\mathbf{h}_k(\mathbf{t}[n])=\mathbf{B}[n]\hat{\mathbf{h}}_k.$
Next, the received signal at user $k$, $k\in\mathcal{K}$, can be rewritten as follows,
\begin{equation}
    {y}_k[n]=\hat{\mathbf{h}}_k^H\mathbf{B}^T[n]\mathbf{x}[n]+{n}_k.
\end{equation}
Thus, the signal-to-interference-plus-noise ratio (SINR) $\mathrm{SINR}_k[n]$ of user $k$ in snapshot $n$ is given by
\begin{equation}
    \begin{aligned}
    \mathrm{SINR}_k[n]=\frac{{\mathrm{Tr}}(\mathbf{B}^T[n]\mathbf{W}_k[n]\mathbf{B}[n]\hat{\mathbf{H}}_k)}{{\mathrm{Tr}}(\mathbf{B}^T[n](\underset{k'\neq k}{\sum}\mathbf{W}_{k'}[n]+\mathbf{R}[n])\mathbf{B}[n]\hat{\mathbf{H}}_k)+\sigma_{k}^2},\notag
    \end{aligned}
\end{equation}
where $\mathbf{R}[n]$ denotes a submatrix of ${\mathbf{R}}$ formed by rows $\{(n-1)M+1,\cdots,nM\}$ and columns $\{(n-1)M+1,\cdots,nM\}$, $\mathbf{W}_k[n]=\mathbf{w}_k[n]\mathbf{w}_k^H[n]$, and  $\hat{\mathbf{H}}_k=\hat{\mathbf{h}}_k\hat{\mathbf{h}}^H_k$. For communication system design, we aim to guarantee a minimum required SINR for all users in all $N$ snapshots, which is given by
\begin{equation}
    \begin{aligned}
    \mbox{C1}: {\mathrm{Tr}}(\mathbf{B}^T[n]\overline{\mathbf{W}}_k[n]\mathbf{B}[n]\hat{\mathbf{H}}_k)-\gamma_k\sigma_k^2\geq 0,\forall k, n,\notag
    \end{aligned}
\end{equation}
where $\overline{\mathbf{W}}_k[n]=\mathbf{W}_k[n]-\gamma_k\sum_{k'\in\mathcal{K}\setminus\{k\}}\mathbf{W}_{k'}[n]-\gamma_k\mathbf{R}[n]$
Next, we define matrix $\mathbf{P}=[\mathbf{p}_1,\cdots,\mathbf{p}_J]$ to collect all $J$ possible discrete positions of the MA elements. Thus, position $\mathbf{t}_m[n]$ of MA element $m$ in snapshot $n$ is given by $\mathbf{t}_m[n]=\mathbf{P}\mathbf{b}_m[n]$.
Hence, we can rewrite the MA motion constraints in \eqref{Max_dis_cons} and \eqref{Min_dis_cons} as 
\begin{equation}
\begin{aligned}
     \mbox{C2}:&\|\mathbf{P}\mathbf{b}_m[n]-\mathbf{P}\mathbf{b}_m[n-1]\|_{\infty}\leq D_{\mathrm{max}},\forall m,n,\,\text{and}\\
     \mbox{C3}:&\|\mathbf{P}\mathbf{b}_m[n]-\mathbf{P}\mathbf{b}_{m'}[n]\|_2\geq D_{\mathrm{min}},\forall n, m'\neq m,
\end{aligned}
\end{equation}
respectively. From the sensing perspective, for a given elevation AoD $\alpha_t$ and azimuth AoD $\beta_t$, we define a vector $\mathbf{a}^H(\alpha_t,\beta_t)=[{a}_{1}(\alpha_t,\beta_t).\cdots,{a}_{J}(\alpha_t,\beta_t)]$ as the virtual steering vector of the transmit area with ${a}_{j}(\alpha_t,\beta_t)=e^{j\rho(\alpha_t,\beta_t,\mathbf{p}_j)}$. Then, the corresponding transmit steering vector in snapshot $n$ is given by $\mathbf{a}_t^H(\alpha_t,\beta_t,\mathbf{t}[n])=\mathbf{a}^H(\alpha_t,\beta_t)\mathbf{B}^T[n]$ and the beam gain can be rewritten as follows,
\begin{equation}
\begin{aligned}
    P(\alpha_t,\beta_t)&=\sigma_s^2\hspace{-3mm}\sum_{n\in\mathcal{N},k\in\mathcal{K}}\hspace{-3mm}\mathbf{a}^H(\alpha_t,\beta_t)\mathbf{B}^T[n]\mathbf{W}_k[n]\mathbf{B}[n]\mathbf{a}(\alpha_t,\beta_t)\\[-3pt]
    &+\mathbf{a}^H(\alpha_t,\beta_t)\mathbf{B}^T\mathbf{R}\mathbf{B}\mathbf{a}(\alpha_t,\beta_t),\notag
\end{aligned}
\end{equation}
where $\mathbf{B}^T=[\mathbf{B}[1]^T,\cdots,\mathbf{B}[N]^T]$.
Moreover, effective target detection requires the DFRC BS to illuminate the desired target using highly directional beams with suppressed sidelobes, such that the echo from the desired target can be distinguished from clutter generated by unintended targets.
To this end, we adopt beam pattern shaping as the main sensing performance metric. Specifically, the angular domain is discretized into $I$ 2D areas. Then, the ideal transmit beam pattern $\{\widetilde{P}(\alpha_i,\beta_i)\}_{i=1}^{I}$ is defined as
\begin{equation}
\widetilde{P}(\alpha_i,\beta_i) =
\begin{cases}
1, & \left| \alpha_i - \alpha_{\mathrm{des}} \right| \le \dfrac{\psi}{2},\,\text{and}\, \left| \beta_i - \beta_{\mathrm{des}} \right| \le \dfrac{\varphi}{2},\\
0, & \text{otherwise},\notag
\end{cases}
\label{eq:ideal_tx_pattern}
\end{equation}
where $\psi$ and $\varphi$ denote the desired beamwidth and $\alpha_{\mathrm{des}}$ and $\beta_{\mathrm{des}}$ are the central angles of the slice.
Next, to quantify the deviation between the ideal beam pattern and the synthesized beam pattern, we define the beam pattern mismatch error as $\mathcal{C}\!\left(\eta, \mathbf{W},\mathbf{R},\mathbf{B}\right)
\triangleq
\sum_{i=1}^{I}
\left|
\eta \widetilde{P}(\alpha_i,\beta_i)
-
P(\alpha_i,\beta_i)
\right|,$
where $\eta \in \mathbb{R}$ is an auxiliary variable, which reflects the average in-slice beam gain and $\mathbf{W}=[\mathbf{W}_1[1],\cdots,\mathbf{W}_K[N]]$. In the paper, we aim to minimize the beam pattern mismatch error to enable high-quality sensing, while satisfying the QoS requirements of the communication users. In particular, the optimal MA trajectory and beamforming vectors can be obtained by solving the following optimization problem
    \begin{align}
        &\underset{\eta,\mathbf{B},\mathbf{W}_k[n]\succeq \mathbf{0},\forall k,n, \mathbf{R}\succeq \mathbf{0}}{\mino}\quad \mathcal{C}\!\left(\eta, \mathbf{W},\mathbf{R},\mathbf{B}[n]\right)\notag\\[-3pt]
        &\mathrm{s.t.}\quad \mbox{C1},\mbox{C2},\mbox{C3},\mbox{C4:}\hspace*{1mm} b_{m,j}[n]\in \{0,1\},\hspace*{1mm} \forall n, m, j,\label{SDR_problem}\\[-2pt]
    &\mbox{C5:}\hspace*{1mm} \sum_{j\in\mathcal{J}}b_{m,j}[n]=1,\hspace*{1mm}\forall m, n,\,\mbox{C6:}\hspace*{1mm} \mathrm{rank}(\mathbf{W}_k[n])=1,\forall k,n \notag\\[-2pt]
    &\mbox{C7:}\hspace*{1mm} \sigma_s^2\sum_{n,k}\mathrm{Tr}(\mathbf{W}_k[n])+\mathrm{Tr}(\mathbf{R})\leq P_{\mathrm{max}},\notag 
    \end{align}
Here, constraints C4 and C5 are imposed due to the binary nature of $\mathbf{B}[n]$. In addition, constraint C7 guarantees that the total power consumption remains under a given threshold $P_{\mathrm{max}}$. Problem \eqref{SDR_problem} is non-convex due to the discrete feasible set of $\mathbf{B}[n]$. 
However, we note that problem \eqref{SDR_problem} is convex for fixed $\mathbf{B}[n]$ if rank constraint C6 is dropped.
Exploiting this property, we develop a globally optimal algorithm based on the BnB method in the following.
\section{Algorithm Design}
\begin{figure}
    \centering
    \includegraphics[width=1.8 in]{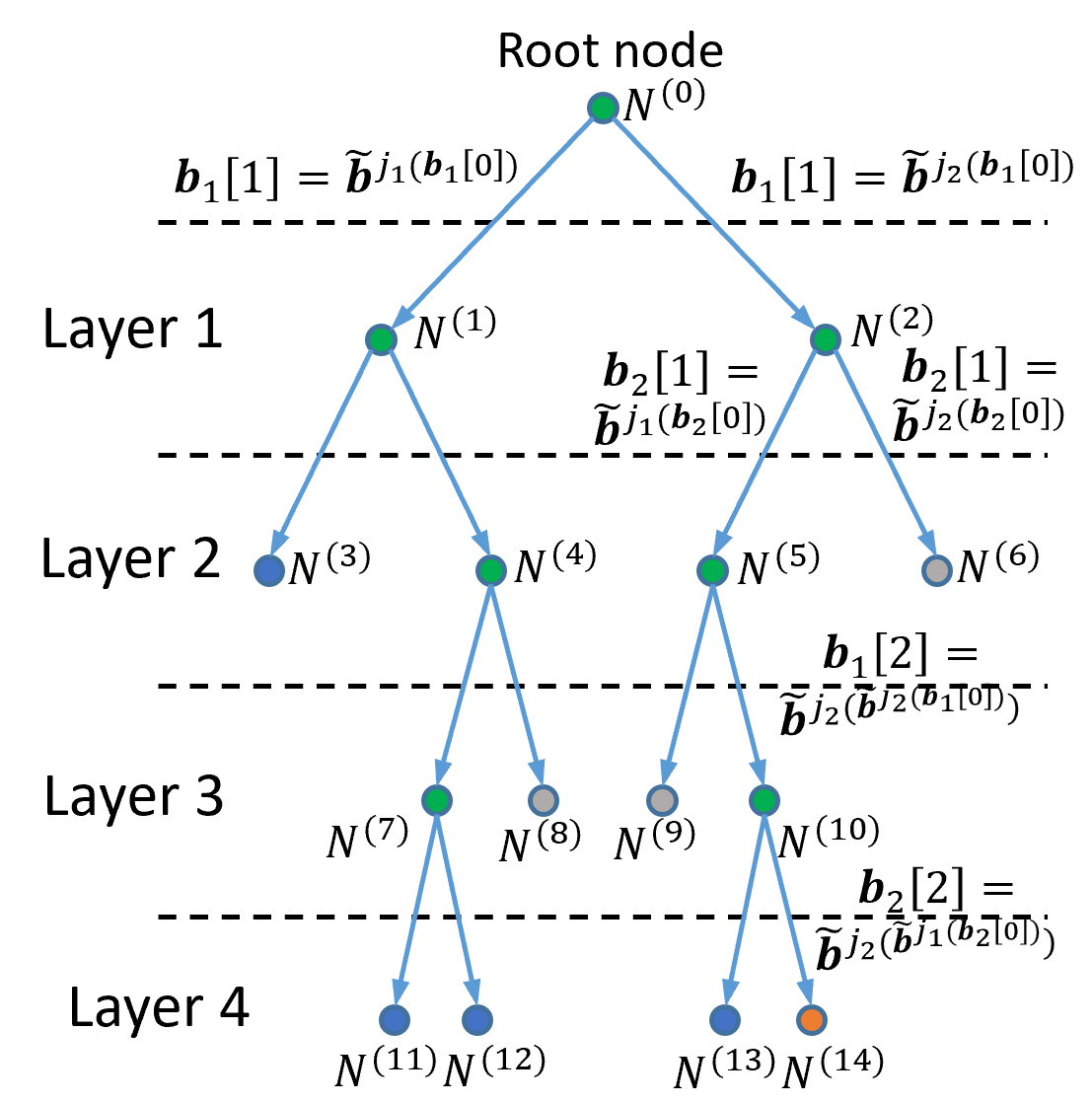}
    \caption{Illustration of the BnB search tree structure for $M = 2$, $N = 2$ and $N_f=2$ in the $7$-th iteration. Green, blue, grey, and orange nodes correspond to internal, external, discarded, and optimal nodes, respectively.}\vspace{1mm}
    \label{fig:BnB}
\end{figure}

Due to the maximum distance constraint C2, in snapshot $n$, each MA element can only select a limited number of candidate positions in the vicinity of its position in snapshot $n-1$. Therefore, once the position of MA element $m$ in snapshot $n-1$ is determined, the feasible set of $\mathbf{b}_m[n]$ is given by a reduced candidate set $\mathcal{B}_{m,n}(\mathbf{b}_m[n-1])=\{\widetilde{\mathbf{b}}^{j_{1}(\mathbf{b}_m[n-1])},\cdots, \widetilde{\mathbf{b}}^{j_{N_f}(\mathbf{b}_m[n-1])}\}$ containing $N_f$ elements, where $\widetilde{\mathbf{b}}^{j_{n_f}(\mathbf{b}_m[n-1])}$ denotes a binary vector with only the $j_{n_f}(\mathbf{b}_m[n-1])$-th  element equal to one, satisfying $\|\mathbf{P}\widetilde{\mathbf{b}}^{j_{n_f}(\mathbf{b}_m[n-1])}-\mathbf{P}\mathbf{b}_m[n-1]\|_{\infty}\leq D_{\mathrm{max}}$. By exploiting this MA mobility structure, we construct a BnB search tree with respect to vectors $\mathbf{b}_m[n]$ to reduce the search space size, as shown in Fig. \ref{fig:BnB}. Specifically, the first $M$ layers correspond to the vectors $\mathbf{b}_1[1],\cdots,\mathbf{b}_M[1]$ in the first snapshot. The layers associated with the subsequent snapshots are then added iteratively.
Then, we introduce the key notations. Let $\mathcal{T}^{(t)}$ denote the BnB search tree at the $t$-th iteration of the algorithm. Next, $\mathcal{N}_i=\{\mathbf{B}^{(i)},l^{(i)},F_L^{(i)},F_U^{(i)}\}$, $\forall i \in\{0,\cdots, N_ft\}$, denotes the $i$-th node of search tree $\mathcal{T}^{(t)}$, where $\mathbf{B}^{(i)}=[\mathbf{b}^{(i)}_1[1],\cdots,\mathbf{b}^{(i)}_{m^{(i)}}[n^{(i)}]]$ collects the already determined binary vector of node $\mathcal{N}_{i}$ and $(m^{(i)},n^{(i)})$ is the index of the last determined binary vector. Then, $l^{(i)}$, $F_L^{(i)}$, and $F_U^{(i)}$ denote the layer index, the lower bound and upper bound corresponding to the node $\mathcal{N}_{i}$, respectively. According to the tree structure shown in Fig. \ref{fig:BnB}, the layer index satisfies the relation
$l^{(i)} = (n^{(i)}-1) \times M + m^{(i)}$.

\subsection{Initialization}


To obtain an initial lower bound for \eqref{SDR_problem}, we first relax the maximum distance constraint C2 to decouple the dependence between variables $\mathbf{b}_m[n]$ and $\mathbf{b}_m[n-1]$. Then, in snapshot $n>1$, we introduce the following relaxed constraint,
\begin{equation}
    \overline{\mbox{C2}}: \|\mathbf{P}\mathbf{b}_m[n]-\mathbf{P}\mathbf{b}_m[0]\|_{\infty}\leq nD_{\mathrm{max}}, \forall m.
\end{equation}
Here, we define a set $\mathcal{I}_{m}[n]$ to collect the indices of $i_f$ such that $\mathbf{b}_m[n]$ satisfies constraint $\overline{\mbox{C2}}$ with ${b}_{m,i_f}[n]=1$ and ${b}_{m,j}[n]=0, j\neq i_f$. Then, to decouple $\mathbf{B}[n]$ and $\mathbf{W}_k[n]$, we define an auxiliary radar covariance variable $\hat{\mathbf{W}}_k[n]=\mathbf{B}^T[n]\mathbf{W}_k[n]\mathbf{B}[n]\succeq\mathbf{0}$ with the following properties \cite{shrestha2023optimal}:
\begin{equation}
\begin{aligned}
    \mbox{C8a}:&\hat{\mathbf{W}}_k[n](i_f,:)=0,\, \hat{\mathbf{W}}_k[n](:,i_f)=0,  i_f\notin\mathcal{I}[n]\\[-2pt]
    \mbox{C8b}:&\|\hat{\mathbf{W}}_k[n]\|_{\mathrm{row-0}}=M,\|\hat{\mathbf{W}}_k[n]\|_{\mathrm{column-0}}=M,
    \\[-2pt]
    \overline{\mbox{C6}}:&\hspace*{1mm} \mathrm{rank}(\hat{\mathbf{W}}_k[n])=1, \hspace*{1mm}\forall k\in \mathcal{K}, n\in\mathcal{N},
\end{aligned}
\end{equation}
where $\mathcal{I}[n]$ denotes the union set of $\mathcal{I}_{1}[n],\cdots, \mathcal{I}_{M}[n]$. Similarly, we define an auxiliary variable $\hat{\mathbf{R}}=\mathbf{B}^T\mathbf{R}\mathbf{B}$, which satisfies the following constraints
\begin{equation}
\begin{aligned}
    \mbox{C8c}:&\hat{\mathbf{R}}(i_f,:)=0,\, \hat{\mathbf{R}}(:,i_f)=0,  i_f\notin\mathcal{I}\\[-2pt]
    \mbox{C8d}:&\|\hat{\mathbf{R}}\|_{\mathrm{row-0}}=N M,\|\hat{\mathbf{R}}\|_{\mathrm{column-0}}=N M,
\end{aligned}
\end{equation}
where $\mathcal{I}$ denotes the union set of $\mathcal{I}[1],\cdots, \mathcal{I}[N]$. Here, by introducing auxiliary variables $\hat{\mathbf{W}}_k[n]$ and $\hat{\mathbf{R}}$, we equivalently recast constraints C2, C4, and C5 as constraints C8a, C8b, C8c, and C8d \cite{shrestha2023optimal}. Then, we can rewrite the objective function, ${\mbox{C1}}$, and ${\mbox{C7}}$ as follows:
    \begin{align}\vspace{-2mm}
        &\mathcal{C}(\eta, \hat{\mathbf{W}}_k[n],\hat{\mathbf{R}},)=\sum_{i=1}^{I}
|
\eta \widetilde{P}(\alpha_i,\beta_i)
-\hat{P}(\alpha_i,\beta_i)
|,\label{C1_reform}\\[-2pt]
&\overline{\mbox{C1}}:\mathrm{Tr}({\hat{\mathbf{H}}_k^H\hat{\overline{\mathbf{W}}}_k[n]})-\gamma_k\sigma_n^2\geq 0,\notag\\[-2pt]
&\overline{\mbox{C7}}: \sigma_s^2\sum_{n\in\mathcal{N}^t}\sum_{k\in\mathcal{K}}\mathrm{Tr}(\hat{\mathbf{W}}_k[n])+\mathrm{Tr}(\hat{\mathbf{R}})\leq P_{\mathrm{max}}\notag
    \end{align}
where $\hat{P}(\alpha_i,\beta_i)=\mathrm{Tr}(\sum_{n\in\mathcal{N}^t}\sum_{k\in\mathcal{K}}\hat{\mathbf{W}}_k[n]\mathbf{A}_T(\alpha_i,\beta_i))+\mathrm{Tr}(\hat{\mathbf{R}}\mathbf{A}_T(\alpha_i,\beta_i)))$,  $\hat{\overline{\mathbf{W}}}_k[n]=\hat{{\mathbf{W}}}_k[n]-\gamma_k{(\sum_{k'\in\mathcal{K}\setminus\{k\}}\hat{\mathbf{W}}_{k'}[n]+\hat{\mathbf{R}}[n]})$ and $\hat{\mathbf{R}}[n]$ denotes the submatrix of $\hat{\mathbf{R}}$ formed by rows $\{(n-1)J+1,\cdots,nJ\}$ and columns $\{(n-1)J+1,\cdots,nJ\}$ and $\mathbf{A}_T(\alpha_t,\beta_t)=\mathbf{a}_T(\alpha_t,\beta_t)\mathbf{a}_T^H(\alpha_t,\beta_t)$. Note that the objective function and constraints $\overline{\mbox{C1}}$ and $\overline{\mbox{C7}}$ are convex with respect to (w.r.t.) $\hat{\mathbf{W}}_k[n]$ and $\hat{\mathbf{R}}$.
Then, by relaxing the minimum distance constraint C3, the number of non-zero row and column constraints C8b, C8d, and the rank-one constraint $\overline{\mbox{C6}}$, a relaxed version of problem \eqref{SDR_problem} is given by
\begin{equation}
\label{Ini_UB_problem}
      \begin{aligned}
        \underset{\eta,\hat{\mathbf{W}}_k[n]\succeq\mathbf{0},\hat{\mathbf{R}}\succeq\mathbf{0}}{\mino}&\quad \mathcal{C}\!\left(\eta, \hat{\mathbf{W}}_k[n],\hat{\mathbf{R}}\right)\\[-8pt]
        \mathrm{s.t.}&\quad \overline{\mbox{C1}},\overline{\mbox{C7}},\mbox{C8a},\mbox{C8c},
    \end{aligned}
\end{equation}
which is convex and can be optimally solved by CVX \cite{grant2008cvx}. Since constraints C2, C4, and C5 are equivalently recast as constraints C8a - C8d, and constraint C3 is ignored, the binary position selection variable matrix $\mathbf{B}$ can be omitted in the optimization \cite{shrestha2023optimal}. Thus, optimally solving problem \eqref{Ini_UB_problem} leads to an initial lower bound of original problem \eqref{SDR_problem}, which is given by $F_L^{(0)}=\mathcal{C}\!\left(\eta^{(0)}_L, \hat{\mathbf{W}}^{(0)}_{L,k}[n],\hat{\mathbf{R}}_L^{(0)}\right)$.

Next, we derive an initial upper bound for problem \eqref{SDR_problem}. Indeed, an upper bound of \eqref{SDR_problem} can be generated based on any feasible $\mathbf{B}[n]$ satisfying constraints ${\mbox{C2}},{\mbox{C3}},{\mbox{C4}}, \mbox{C5}$ \cite{boyd2007branch}. For problem \eqref{SDR_problem}, a feasible solution of $\mathbf{B}[n]$ is given by $\mathbf{B}^{(0)}[n]=\mathbf{B}[0], \forall n$, which implies that all MA elements stay in the initial position in all $N$ snapshots. Here, the initial position of the MA elements must satisfy the minimum distance constraint C3. Thus, based on the binary solution $\mathbf{B}^{(0)}[n]$, an upper bound for \eqref{SDR_problem} is obtained by solving the optimization problem in \eqref{SDR_problem} and relaxing the rank-one constraint C6 for fixed $\mathbf{B}[n]=\mathbf{B}^{(0)}[n]$, e.g., by using CVX\cite{grant2008cvx}. The tightness of the resulting relaxation can be proven in a similar manner as in \cite[Theorem 1]{xu2022robust}, which is omitted here due to page limitations. Here, $\eta_U^{(0)}$, $\mathbf{W}_{U,k}^{(0)}[n]$, and $\mathbf{R}_U^{(0)}$ denote the optimal $\eta$, $\mathbf{W}_k[n]$, and $\mathbf{R}$ obtained by solving \eqref{SDR_problem} for fixed $\mathbf{B}[n]=\mathbf{B}^{(0)}[n]$. Then, an initial upper bound of \eqref{SDR_problem} is given by $F_U^{(0)}=\mathcal{C}\!\left(\eta^{(0)}_U, \hat{\mathbf{W}}^{(0)}_{U,k}[n],\hat{\mathbf{R}}_U^{(0)}\right)$.

\subsection{Branch and Bound}
At the beginning of the $t$-th iteration, the BnB search tree inherited from the previous iteration is denoted by $\mathcal{T}^{(t-1)}$. Moreover, we define $\mathcal{I}^{(t-1)}$ and $\mathcal{E}^{(t-1)}$ as the sets of internal and external nodes of the tree $\mathcal{T}^{(t-1)}$, respectively, as illustrated in Fig. \ref{fig:BnB}. In this paper, internal and external nodes represent the nodes with and without child nodes, respectively. 
Then, we select the node $\mathcal{N}_{i_t}=\{\mathbf{B}^{(i_t)},l^{(i_t)},F_L^{(i_t)},F_U^{(i_t)}\}$, $\mathcal{N}_{i_t}\in\mathcal{E}^{(t-1)}$, from the external nodes of the BnB tree $\mathcal{T}^{(t-1)}$ with the smallest lower bound $F_L^{(i_t)}$, where $i_t$ denotes the index of the selected node. 
According to the BnB tree structure, we have $l^{(i_t)} = (n^{(i_t)}-1) \times M + m^{(i_t)}$. For node $\mathcal{N}_{i_t}$, the positions of all MA elements in $n\leq n^{(i_t)}=\lfloor l^{(i_t)}/M \rfloor$ snapshots are determined. Moreover, in snapshot $\lceil l^{(i_t)}/M \rceil$, the positions of the MA elements with index from $1$ to $m^{(i_t)}=l^{(i_t)} \mod M$ are determined. Here, we define sets $\mathcal{M}_{i_t}=\{(1,1),\cdots, (n^{(i_t)},m^{(i_t)})\}$ to collect the indices of the determined binary vectors of node $\mathcal{N}_{i_t}$.

Then, we partition the selected node $\mathcal{N}_{i_t}$ into $N_f$ child nodes based on the search order defined by the BnB search tree, as shown in Fig. \ref{fig:BnB}. Here, $m^{(i_t)}$ and $n^{(i_t)}$ are the indices of the last determined binary variable of node $\mathcal{N}_{i_t}$. If $m^{(i_t)}<M$, we select the binary vector $\mathbf{b}_{m^{(i_t)}+1}[n^{(i_t)}]$ for partitioning. Otherwise, the binary vector $\mathbf{b}_{1}[n^{(i_t)}+1]$ is selected. Let $\mathbf{b}_{m^{(t)}}[n^{(t)}]$ denote the binary vector selected for partitioning. Then, we can obtain its feasible set $\mathcal{B}_{m^{(t)},n^{(t)}}(\mathbf{b}_{m^{(t)}}[n^{(t)}-1])$ satisfying constraint C2 via an exhaustive search over $J$ candidate binary vectors.
Then, in the $t$-th iteration, resource allocation problems $\mathcal{P}_i, i\in\{1,\cdots,N_f\}$, corresponding to the $N_f$ child nodes are given by
\begin{eqnarray}
\label{LB_Problem_branch}
    \mathcal{P}_i:&&\hspace*{-4mm}\underset{\eta,\mathbf{B}[n],\mathbf{W}_k[n]\succeq\mathbf{0},\mathbf{R}\succeq\mathbf{0}}{\mino}\quad \mathcal{C}\left(\eta, \mathbf{W}_k[n],\mathbf{R},\mathbf{B}[n]\right)\notag\\[-4pt]
    &&\hspace*{-4mm}\mbox{s.t.}\hspace*{2mm} {\mbox{C1}}-\mbox{C7},\,{\mbox{C9a}}:\mathbf{b}_m[n]=\mathbf{b}^{(i_t)}_m[n],\,\forall (m,n)\in\mathcal{M}_{i_t}\notag\\
    &&\hspace*{2mm} {\mbox{C9b}}:\mathbf{b}_{m^{(t)}}[n^{(t)}]=\widetilde{\mathbf{b}}^{j_{i}(\mathbf{b}_{m^{(t)}}[n^{(t)}-1])}.
\end{eqnarray}
{If the selected $\mathbf{b}_{m^{(t)}}[n^{(t)}]=\widetilde{\mathbf{b}}^{j_{i}(\mathbf{b}_{m^{(t)}}[n^{(t)}-1])}$ and the resulting binary matrix $\mathbf{B}^{(i_t)}[n]$ violates the minimum-distance constraint C3, the corresponding node is marked as infeasible with lower bound $+\infty$.}
Similar to the procedure for deriving the initial lower bound, we first relax constraint C2 for the undetermined binary vectors $\mathbf{b}_m[n]$. For snapshots $n$, $n\geq n^{(i_t)}$, we define a relaxed constraint $\widetilde{\mbox{C2}}$ as 
\begin{equation}
\begin{aligned}
    &\|\mathbf{P}(\mathbf{b}_m[n]-\mathbf{b}_m[n^{(i_t)}])\|_{\infty}\leq (n-n^{(i_t)})D_{\mathrm{max}}, \forall m\leq m^{(i_t)},\\
    &\|\mathbf{P}(\mathbf{b}_m[n]\hspace{-0.5mm}-\hspace{-0.5mm}\mathbf{b}_m[n^{(i_t)}-1])\|_2^2\hspace{-0.5mm}\leq\hspace{-0.5mm} (n\hspace{-0.5mm}-\hspace{-0.5mm}n^{(i_t)}\hspace{-0.5mm}+\hspace{-0.5mm}1)D_{\mathrm{max}}, \text{else}.\notag
\end{aligned}
\end{equation}
Here, we define a set $\widetilde{\mathcal{I}}_{m}[n],n\geq n^{(i_t)}$, to collect the indices of $i_f$ such that $\mathbf{b}_m[n]$ satisfies constraint $\widetilde{\mbox{C2}}$ with ${b}_{m,i_f}[n]=1$ and ${b}_{m,j}[n]=0, j\neq i_f$. Moreover, for $n\leq n^{(i_t)}$, we define $\widetilde{\mathcal{I}}_{m}[n]=\{j^{(i_t)}_m[n]\}$, where $j^{(i_t)}_m[n]$ denote the index of non-zero element in determined binary vector $\mathbf{b}^{(i_t)}_m[n]$. Then, to decouple $\mathbf{B}[n]$ and $\mathbf{W}_k[n]$, we define an auxiliary variable $\widetilde{\mathbf{W}}_k[n]=\mathbf{B}^T[n]\mathbf{W}_k[n]\mathbf{B}[n]$, $n\geq n_{i_t}$, which has the following properties:
\begin{equation}
\begin{aligned}
    \widetilde{\mbox{C8a}}:&\widetilde{\mathbf{W}}_k[n](i_f,:)=0,\, \widetilde{\mathbf{W}}_k[n](:,i_f)=0,  i_f\notin\widetilde{\mathcal{I}}[n]\\
    \widetilde{\mbox{C8b}}:&\|\widetilde{\mathbf{W}}_k[n]\|_{\mathrm{row-0}}=M,\|\widetilde{\mathbf{W}}_k[n]\|_{\mathrm{column-0}}=M,\\
    \widetilde{\mbox{C6}}:&\hspace*{1mm} \mathrm{rank}(\widetilde{\mathbf{W}}_k[n])=1, \hspace*{1mm}\forall k\in \mathcal{K}, n\in\mathcal{N},
\end{aligned}
\end{equation}
where $\widetilde{\mathcal{I}}[n]$ denotes the union set of $\widetilde{\mathcal{I}}_{1}[n],\cdots, \widetilde{\mathcal{I}}_{M}[n]$. Similarly, we define $\widetilde{\mathbf{R}}=\mathbf{B}^T\mathbf{R}\mathbf{B}$ satisfying constraints
\begin{equation}
\begin{aligned}
    \widetilde{\mbox{C8c}}:&\widetilde{\mathbf{R}}(i_f,:)=0,\, \widetilde{\mathbf{R}}(:,i_f)=0,  i_f\notin\mathcal{I}\\
    \widetilde{\mbox{C8d}}:&\|\widetilde{\mathbf{R}}\|_{\mathrm{row-0}}=N M,\|\widetilde{\mathbf{R}}\|_{\mathrm{column-0}}=N M,
\end{aligned}
\end{equation}
where $\widetilde{\mathcal{I}}$ denotes the union set of $\widetilde{\mathcal{I}}[1],\cdots, \widetilde{\mathcal{I}}[N]$. Next, constraints C1 and C7 can be recast into constraints $\widetilde{\mbox{C1}}$ and $\widetilde{\mbox{C7}}$ in a similar manner as \eqref{C1_reform}. In particular, variables $\hat{\mathbf{W}}_k[n]$ and $\hat{\mathbf{R}}$ are replaced by $\widetilde{\mathbf{W}}_k[n]$ and $\widetilde{\mathbf{R}}$, respectively.
Then, a lower bound of \eqref{LB_Problem_branch} is obtained by solving the following relaxed problems $\widetilde{\mathcal{P}}_i$, $i\in\{1,\cdots, N_f\}$:
\begin{eqnarray}
\label{LB_Problem_branch_LB}
    \widetilde{\mathcal{P}}_i:&&\underset{\eta,\widetilde{\mathbf{W}}_k[n],\widetilde{\mathbf{R}}}{\mino}\quad \mathcal{C}\!\left(\eta, \widetilde{\mathbf{W}}_k[n],\widetilde{\mathbf{R}}\right)\\[-6pt]
    &&\hspace*{-4mm}\mbox{s.t.}\hspace*{2mm} \widetilde{\mbox{C1}},\widetilde{\mbox{C7}},\widetilde{\mbox{C8a}},\widetilde{\mbox{C8c}},\mbox{C9a},\mbox{C9b},\notag
\end{eqnarray}
where the non-convex constraints C3, $\widetilde{\mbox{C6}},$ $\widetilde{\mbox{C8b}}$ and $\widetilde{\mbox{C8d}}$ are relaxed.
Note that $\widetilde{\mathcal{P}}_i$ is a convex optimization problem and can be optimally solved by CVX. Here, the optimal solution of relaxed problem $\widetilde{\mathcal{P}}_{i},\forall i\in\{1,\cdots,N_f\}$, is denoted as $(\eta^{(t)}_{L,i},\widetilde{\mathbf{W}}_{L,i,k}^{(t)}[n],\widetilde{\mathbf{R}}_{L,i}^{(t)})$. The lower bound of problem ${\mathcal{P}}_{i},\,\forall i\in\{1,\cdots,N_f\}$, is given by the objective function value $F_{L,i}^{(t)}=\mathcal{C}\!\left(\eta^{(t)}_{L,i},\widetilde{\mathbf{W}}_{L,i,k}^{(t)}[n],\widetilde{\mathbf{R}}_{L,i}^{(t)}\right)$. 

Next, we derive the corresponding upper bounds for the inserted new nodes.
For the $i$-th node, we generate a feasible matrix $\mathbf{B}_{i}^{(t)}[n^{(t)}]$ satisfying constraints C2-C5, C9a, and C9b by exhaustive search. Then, we can generate a feasible $\mathbf{B}_{i}^{(t)}=[\mathbf{B}_{i}^{(t)}[1],\cdots,\mathbf{B}_{i}^{(t)}[N]]$ as follows:
\begin{equation}\label{Feasible_B}
\begin{aligned}
    &\mathbf{B}_{i}^{(t)}[n]=\mathbf{B}^{(i_t)}[n], \forall n<n^{(t)}\\[-3pt]
    &\mathbf{B}_{i}^{(t)}[n]=\mathbf{B}^{(i_t)}[n^{(t)}], \forall n\geq n^{(t)}.
\end{aligned}
\end{equation}
Subsequently, we solve problem \eqref{LB_Problem_branch} by fixing the binary variables according to $\mathbf{B}=\mathbf{B}_{i}^{(t)}$ and obtain the corresponding upper bound $F_{U,i}^{(t)}$. After deriving these lower and upper bounds, we expand the BnB tree $\mathcal{T}^{(t-1)}$ by introducing $N_f$ new child nodes of node $\mathcal{N}_{i_t}$ corresponding to $\mathbf{b}_{m^{(t)}}[n^{(t)}]=\widetilde{\mathbf{b}}^{j_{i}(\mathbf{b}_{m^{(t)}}[n^{(t)}-1])}$. 
Let $\mathcal{T}^{(t)}$ denote the BnB tree after introducing these new child nodes. Then, the global lower and upper bounds in the $t$-th iteration, i.e., $\mathrm{LB}^{(t)}$ and $\mathrm{UB}^{(t)}$, are given by the smallest lower and upper bounds among all external nodes in $\mathcal{T}^{(t)}$.
Next, the search tree is pruned by discarding nodes whose lower bounds are no better than the current upper bound $\mathrm{UB}^{(t)}$, as shown in Fig. \ref{fig:BnB}.
\setlength{\textfloatsep}{0pt}
\begin{algorithm}[t]
\caption{Proposed BnB-based Algorithm}
\begin{algorithmic}[1]
\small
\STATE Solve problem \eqref{Ini_UB_problem} to obtain initial lower bound $\mathrm{LB}=F_L^{(0)}$.
\STATE Compute the initial upper bound $\mathrm{UB}=F_U^{(0)}$ by solving problem in \eqref{SDR_problem} for fixed $\mathbf{B}[n]=\mathbf{B}^{(0)}[n]$.
\STATE Initialize the BnB tree $\mathcal{T}^{(0)}=\{\mathcal{N}_0\}$. Set the convergence tolerance $0\leq\Delta_{\mathrm{BnB}}$ and iteration index $t=0$.
\REPEAT
\STATE $t=t+1$,
\STATE Select the external node $\mathcal{N}_{i_t}$ with smallest lower bound.
\STATE Determine the branching index $(m^{(t)},n^{(t)})$. Divide the search space of $\mathcal{N}_{i_t}$ into $N_f$ subspaces corresponding to $\mathbf{b}_{m^{(t)}}[n^{(t)}]=\widetilde{\mathbf{b}}^{j_{i}(\mathbf{b}_{m^{(t)}}[n^{(t)}-1])}$,$\forall i =1,\cdots,N_f$.    
\STATE Solve the relaxed version of the $N_f$ subproblems $\mathcal{P}_{i},$ $i=1,\cdots,N_f$, in \eqref{LB_Problem_branch_LB} to obtain the corresponding lower bounds $F_{L,i}^{(t)}$, $i=1,\cdots,N_f$.
\STATE Compute $\mathbf{B}_{i}^{(t)}$, $i=1,\cdots,N_f$, according to \eqref{Feasible_B}.
\STATE Solve problem in \eqref{LB_Problem_branch} by fixing $\mathbf{B}=\mathbf{B}_{i}^{(t)}$, $i=1,\cdots,N_f$. Store the upper bounds $F_{U,i}^{(t)}$, $i=1,\cdots,N_f$.
\STATE Expand the tree by adding the $N_f$ child nodes corresponding to $\mathbf{b}_{m^{(t)}}[n^{(t)}]=\widetilde{\mathbf{b}}^{j_{i}(\mathbf{b}_{m^{(t)}}[n^{(t)}-1])}$,$\forall i =1,\cdots,N_f$, for the selected node $\mathcal{N}_{i_t}$. 
\STATE Update $\mathrm{LB}^{(t)}$ and $\mathrm{UB}^{(t)}$ as the smallest lower bound and upper bound among the external nodes in the tree.
\UNTIL $\mathrm{UB}^{(t)}-\mathrm{LB}^{(t)}\leq \Delta_{\mathrm{BnB}}$
\end{algorithmic}
\end{algorithm}
\subsection{Overall Algorithm}
The overall procedure of the proposed BnB method is summarized in Algorithm 1. According to the theoretical results in \cite{boyd2007branch}, the proposed BnB-based framework is guaranteed to terminate after a finite number of iterations and achieves an $\epsilon$-optimal solution for any prescribed convergence tolerance $\Delta_{\mathrm{BnB}}\geq 0$. 
Moreover, our simulation results indicate that the proposed algorithm converges within significantly fewer iterations than would be required for an exhaustive search.
 \section{Numerical Results}
We study an MA-assisted MISO downlink system, where the BS employs $M=4$ MA elements to simultaneously serve $K=3$ single-antenna users. The operating carrier frequency is fixed at $28$ GHz. The transmit region at the BS is a square area of size $l\lambda \times l\lambda$, where $l=4$ represents the normalized size of the transmit region. 
The minimum inter-MA-distance is set to $D_{\mathrm{min}} = 5$ mm. The user–BS distances are randomly generated following a uniform distribution between $10$ m and $50$ m. The noise power at each user terminal is assumed to be $-80$ dBm for all $k \in \mathcal{K}$.
 The effective channel between the BS and the $k$-th user $\hat{\mathbf{h}}_k$ is modelled as Rician distributed and modeled as
 $   \hat{\mathbf{h}}_k=\sqrt{L_0D_k^{-\alpha}}\left(\sqrt{\frac{\beta}{1+\beta}}\mathbf{h}^{\mathrm{L}}_k+\sqrt{\frac{1}{1+\beta}}\mathbf{h}^{\mathrm{N}}_k\right),$
where $L_0$ denotes the large-scale fading at reference distance $d_0=1$ m. $\alpha=2.2$ and $\beta=4$ denote the path loss exponent and the Rician factor, respectively. Vectors $\mathbf{h}^{\mathrm{L}}_k$ and $\mathbf{h}^{\mathrm{N}}_k$ denote the LoS component and the NLoS component, respectively, where $\mathbf{h}^{\mathrm{L}}_k$ is modeled by the array response vector of an effective antenna array with $J$ antenna elements, where the $j$-th antenna element is located at $\mathbf{p}_j$. Then, $\mathbf{h}^{\mathrm{N}}_k$ is generated according to a Rayleigh distribution.
 In this work, the initial positions $\mathbf{t}[0]=\mathbf{P}\mathbf{B}[0]$ of the MA elements are randomly generated while satisfying constraint C3. Moreover, the time duration of the MA motion and ISAC slots is $T_{\mathrm{MA}}=10$ ms and $T_{\mathrm{Data}}=90$ ms, respectively. 
In addition, 
the maximum speed and step size of MA motion are given by $v_{\mathrm{MA}}= 0.4$ $\mathrm{mm}/\mathrm{ms}$ and $ 2$ mm, respectively.
Thus, the maximal moving distance of an MA is given by $D_{\mathrm{max}}=v_{\mathrm{MA}}T_{\mathrm{MA}}=4$ mm. Without loss of generality, we assume all users impose an identical SINR requirement, i.e., $\gamma_k=\gamma=10\,\mathrm{dB}, \forall k$. The convergence tolerances $\Delta_{\mathrm{BnB}}$ is set as $\Delta_{\mathrm{BnB}}=10^{-4}$. 

We consider two baseline schemes for comparison. For baseline scheme 1, in $N$ snapshots, the MA elements are kept fixed at $M$ positions, i.e., we have binary matrices $\mathbf{B}[N]=\cdots=\mathbf{B}[1]=\mathbf{B}[0]$. The beamforming matrices $\mathbf{W}_k[n]$ and $\mathbf{R}$ are obtained by solving problem 
\eqref{SDR_problem} for the fixed matrices $\mathbf{B}[N]=\cdots=\mathbf{B}[1]=\mathbf{B}[0]$. For baseline scheme 2, we adopt a random MA movement strategy. Specifically, $\mathbf{B}_{\mathrm{rdn}}$ is randomly generated subject to constraints C2, C3, C4, and C5. Then, the beamforming matrices $\mathbf{W}_k[n]$ and $\mathbf{R}$ are optimized by solving the problem in \eqref{SDR_problem} with fixed $\mathbf{B}=\mathbf{B}_{\mathrm{rdn}}$.
\begin{figure}[htbp]\vspace*{-8mm}
	\centering
    \begin{minipage}{1.68 in}
                 \centering
    \includegraphics[width=1.65 in]{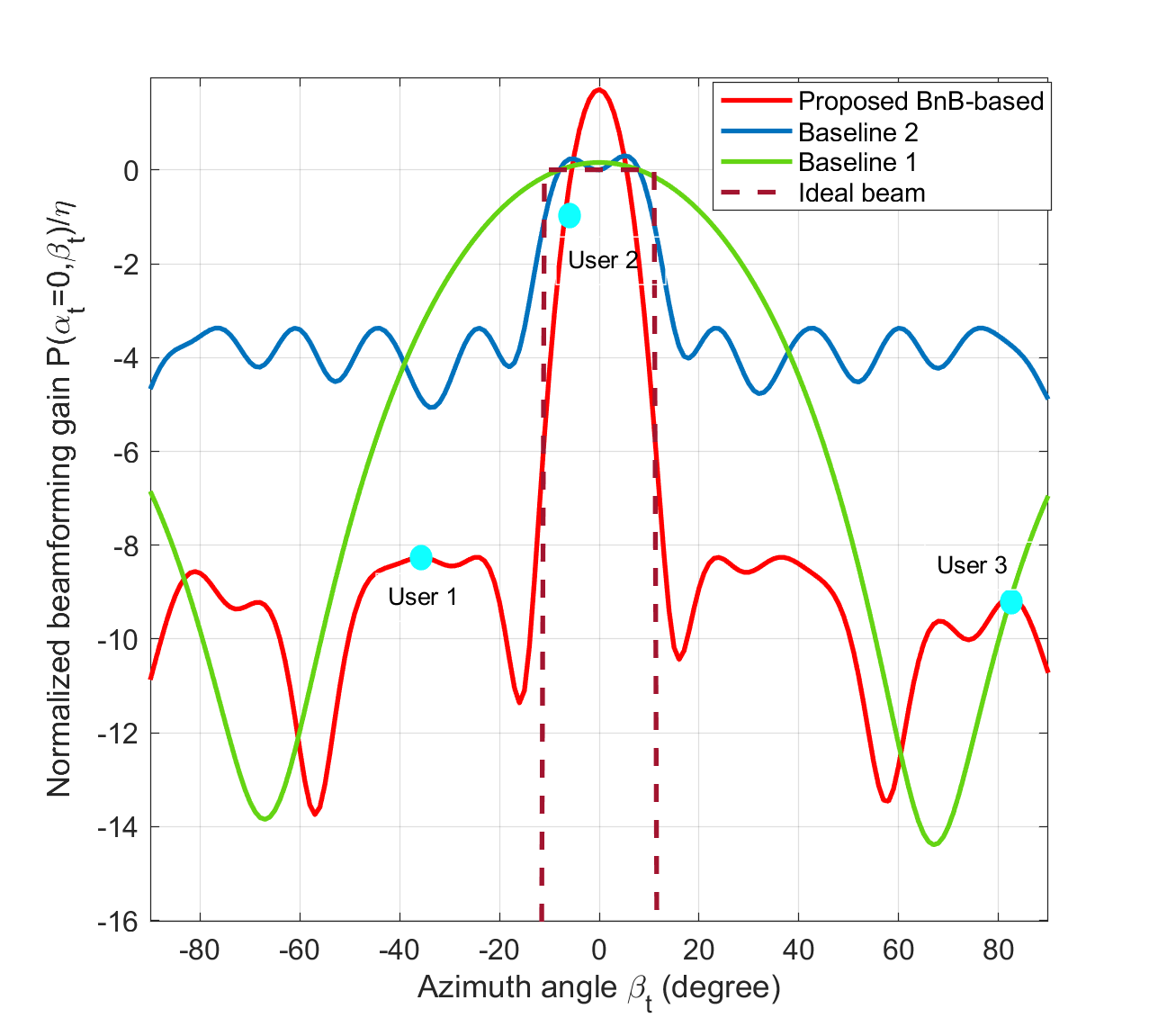}
    \caption{Illustration of normalized beam gain of different schemes with $N=3$, $\alpha_{\mathrm{des}}=0$, $\beta_{\mathrm{des}}=0$, $\psi=0$, and $\varphi=\pi/16$.}
    \label{fig:Beam_pattern}
        
    \end{minipage}
    \begin{minipage}{1.68 in}\vspace{0mm}
                  \centering
    \includegraphics[width=1.65 in]{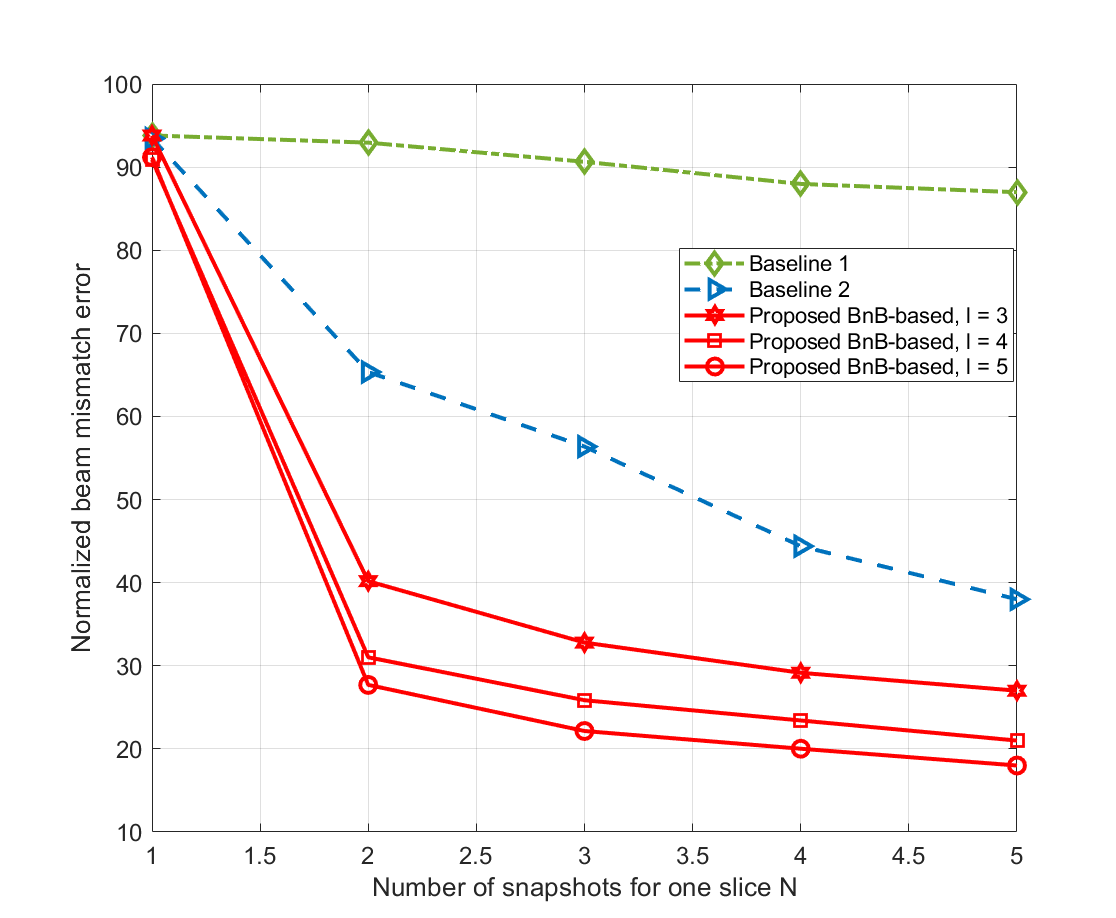}
    \caption{Illustration of average normalized beam mismatch error within the desired region for different schemes and different normalized size of transmit region $l$ with $\alpha_{\mathrm{des}}=0$, $\beta_{\mathrm{des}}=0$, $\psi=\pi/8$, and $\varphi=\pi/8$}
    \label{fig:eta}
    \end{minipage}
\end{figure}\vspace{-5mm}

Fig. \ref{fig:Beam_pattern} illustrates the beam patterns produced by different schemes for a representative channel realization. For ease of visualization, the elevation beamwidth $\psi$ is set to zero, and the beam gain is normalized as $P(0,\beta_t)/\eta$. As can be observed, the beam pattern generated by the proposed BnB-based method achieves a high main-lobe gain in the desired direction while effectively suppressing sidelobes, closely matching the characteristics of the ideal sensing beam pattern. Specifically, by jointly optimizing the MA trajectories and BS ISAC transmit signals, the signals radiated from different MA elements at different snapshots are aligned in phase toward the desired sensing direction. This adaptive co-phased transmission enables constructive superposition of the reflected signals at the desired angle within the sensing angular slice, thereby producing a high main-lobe gain pointing into the desired direction. Besides the main lobe, several pronounced sidelobes are present. In fact, the BS must allocate a portion of the transmit power to serve the communication users. In contrast, baseline scheme 1 exhibits a wider main lobe due to the limited aperture of the fixed antenna array, while baseline scheme 2 shows insufficient sidelobe suppression, underscoring the importance of MA trajectory optimization for improving angular selectivity and interference suppression in ISAC systems.

Fig. \ref{fig:eta} depicts the average normalized beam mismatch error as a function of the number of snapshots $N$ allocated per slice for different schemes. To ensure a fair comparison, the beam mismatch error $\mathcal{C}\!\left(\eta, \mathbf{W}_k[n],\mathbf{R},\mathbf{B}[n]\right)$ is normalized by 
$\eta$. In particular, increasing $N$ monotonically reduces the average beam mismatch error for all schemes, since a larger number of snapshots enables more effective spatial combining. Moreover, the performance gap between baseline scheme 1 and the proposed design widens with increasing $N$, as additional snapshots provide more spatial sampling points for MA movement and formulate a larger virtual aperture, which cannot be exploited by fixed-antenna systems. 
Notably, even with a small snapshot budget (e.g., $N=2, 3$), the proposed scheme achieves low mismatch error. In practice, the MA-enabled ISAC transmitter can operate reliably with only two or three sensing snapshots per slice to reduce the sensing overhead and the computational complexity of the proposed framework. 
\section{Conclusion}
This paper studied a dynamic MA-enhanced ISAC system and investigated the joint design of MA trajectories and beamforming to improve sensing performance, while satisfying the QoS requirements of communication users. The resulting joint optimization problem was formulated as an MINLP problem and then optimally solved using a BnB-based method. Numerical results demonstrated significant sensing performance gain of the proposed scheme over several baseline schemes at the expense of one or two steps of additional antenna repositioning, suggesting a trade-off between ISAC performance and hardware complexity. These findings establish antenna mobility as a new physical-layer design dimension for ISAC systems and motivate further research on low-complexity implementations.
\bibliographystyle{IEEEtran}
\bibliography{reference.bib}
\end{document}